\documentstyle[12pt,psfig]{article}
\begin{document}
\def\e{\mbox{e}}
\def\sgn{{\rm sgn}}
\def\gsim{\;\raise0.3ex\hbox{$>$\kern-0.75em\raise-1.1ex\hbox{$\sim$}}\;}
\def\lsim{\;\raise0.3ex\hbox{$<$\kern-0.75em\raise-1.1ex\hbox{$\sim$}}\;}
\def\MeV{\rm MeV}
\def\eV{\rm eV}
\newcommand {\ignore}[1]{}
\newcommand{\nota}[1]{\makebox[0pt]{\,\,\,\,\,/}#1}
\newcommand{\notp}[1]{\makebox[0pt]{\,\,\,\,/}#1}
\newcommand{\braket}[1]{\mbox{$<$}#1\mbox{$>$}}
\newcommand{\Frac}[2]{\frac{\displaystyle #1}{\displaystyle #2}}
\renewcommand{\arraystretch}{1.5}
\newcommand{\noi}{\noindent}
\newcommand{\bc}{\begin{center}}
\newcommand{\ec}{\end{center}}
\newcommand{\epm}{e^+e^-}
\def\ifmath#1{\relax\ifmmode #1\else $#1$\fi}
\def\half{\ifmath{{\textstyle{1 \over 2}}}}
\def\quarter{\ifmath{{\textstyle{1 \over 4}}}}
\def\3quarter{{\textstyle{3 \over 4}}}
\def\third{\ifmath{{\textstyle{1 \over 3}}}}
\def\twothirds{{\textstyle{2 \over 3}}}
\def\fourth{\ifmath{{\textstyle{1\over 4}}}}
\def\sqrthalf{\ifmath{{\textstyle{1\over\sqrt2}}}}
\def\halfsqrthalf{\ifmath{{\textstyle{1\over2\sqrt2}}}}
\def\cl{\centerline}
\def\vs{\vskip}
\def\hs{\hskip}
\def\ra{\rightarrow}
\def\Ra{\Rightarrow}
\def\us{\undertext}
\overfullrule 0pt
\def\lf{\leaders\hbox to 1em{\hss.\hss}\hfill}
\def\21{$SU(2) \ot U(1)$}
\def\321{$SU(3) \ot SU(2) \ot U(1)$}
\def\ne{\hbox{$\nu_e$ }}
\def\nm{\hbox{$\nu_\mu$ }}
\def\nt{\hbox{$\nu_\tau$ }}
\def\ns{\hbox{$\nu_{s}$ }}
\def\nx{\hbox{$\nu_x$ }}
\def\Nt{\hbox{$N_\tau$ }}
\def\nr{\hbox{$\nu_R$ }}
\def\O{\hbox{$\cal O$ }}
\def\L{\hbox{$\cal L$ }}
\def\mne{\hbox{$m_{\nu_e}$ }}
\def\mnm{\hbox{$m_{\nu_\mu}$ }}
\def\mnt{\hbox{$m_{\nu_\tau}$ }}
\def\mq{\hbox{$m_{q}$}}
\def\ml{\hbox{$m_{l}$}}
\def\mup{\hbox{$m_{u}$}}
\def\md{\hbox{$m_{d}$}}
%
\def\ie{\hbox{\it i.e., }}        \def\etc{\hbox{\it etc. }}
\def\eg{\hbox{\it e.g., }}        \def\cf{\hbox{\it cf.}}
\def\etal{\hbox{\it et al., }}
\def\neus{\hbox{neutrinos }}
\def\gau{\hbox{gauge }}
\def\neu{\hbox{neutrino }}
\def\tr{\mathop{\rm tr}}
\def\Tr{\mathop{\rm Tr}}
\def\Im{\mathop{\rm Im}}
\def\Re{\mathop{\rm Re}}
\def\bR{\mathop{\bf R}}
\def\bC{\mathop{\bf C}}
\def\eq#1{{eq. (\ref{#1})}}
\def\Eq#1{{Eq. (\ref{#1})}}
\def\Eqs#1#2{{Eqs. (\ref{#1}) and (\ref{#2})}}
\def\Eqs#1#2#3{{Eqs. (\ref{#1}), (\ref{#2}) and (\ref{#3})}}
\def\Eqs#1#2#3#4{{Eqs. (\ref{#1}), (\ref{#2}), (\ref{#3}) and (\ref{#4})}}
\def\eqs#1#2{{eqs. (\ref{#1}) and (\ref{#2})}}
\def\eqs#1#2#3{{eqs. (\ref{#1}), (\ref{#2}) and (\ref{#3})}}
\def\eqs#1#2#3#4{{eqs. (\ref{#1}), (\ref{#2}), (\ref{#3}) and (\ref{#4})}}
\def\fig#1{{Fig. (\ref{#1})}}
\def\partder#1#2{{\partial #1\over\partial #2}}
\def\secder#1#2#3{{\partial^2 #1\over\partial #2 \partial #3}}
\def\bra#1{\left\langle #1\right|}
\def\ket#1{\left| #1\right\rangle}
\def\VEV#1{\left\langle #1\right\rangle}
\let\vev\VEV
\def\gdot#1{\rlap{$#1$}/}
\def\abs#1{\left| #1\right|}
\def\pri#1{#1^\prime}
\def\ltap{\raisebox{-.4ex}{\rlap{$\sim$}} \raisebox{.4ex}{$<$}}
\def\gtap{\raisebox{-.4ex}{\rlap{$\sim$}} \raisebox{.4ex}{$>$}}
\def\lsim{\raise0.3ex\hbox{$\;<$\kern-0.75em\raise-1.1ex\hbox{$\sim\;$}}}
\def\gsim{\raise0.3ex\hbox{$\;>$\kern-0.75em\raise-1.1ex\hbox{$\sim\;$}}}
\def\half{{1\over 2}}
\def\beq{\begin{equation}}
\def\eeq{\end{equation}}
\def\bef{\begin{figure}}
\def\eef{\end{figure}}
\def\bet{\begin{table}}
\def\eet{\end{table}}
\def\bea{\begin{eqnarray}}
\def\ba{\begin{array}}
\def\ea{\end{array}}
\def\bi{\begin{itemize}}
\def\ei{\end{itemize}}
\def\ben{\begin{enumerate}}
\def\een{\end{enumerate}}
\def\ra{\rightarrow}
\def\ot{\otimes}
%
\def\com#1#2{
        \left[#1, #2\right]}
\def\eea{\end{eqnarray}}
\def\bentarrow{\:\raisebox{1.3ex}{\rlap{$\vert$}}\!\rightarrow}
\def\longbent{\:\raisebox{3.5ex}{\rlap{$\vert$}}\raisebox{1.3ex}%
        {\rlap{$\vert$}}\!\rightarrow}
\def\onedk#1#2{
        \begin{equation}
        \begin{array}{l}
         #1 \\
         \bentarrow #2
        \end{array}
        \end{equation}
                }
\def\dk#1#2#3{
        \begin{equation}
        \begin{array}{r c l}
        #1 & \rightarrow & #2 \\
         & & \bentarrow #3
        \end{array}
        \end{equation}
                }
\def\dkp#1#2#3#4{
        \begin{equation}
        \begin{array}{r c l}
        #1 & \rightarrow & #2#3 \\
         & & \phantom{\; #2}\bentarrow #4
        \end{array}
        \end{equation}
                }
\def\bothdk#1#2#3#4#5{
        \begin{equation}
        \begin{array}{r c l}
        #1 & \rightarrow & #2#3 \\
         & & \:\raisebox{1.3ex}{\rlap{$\vert$}}\raisebox{-0.5ex}{$\vert$}%
        \phantom{#2}\!\bentarrow #4 \\
         & & \bentarrow #5
        \end{array}
        \end{equation}
                }
%
%
\def\ap#1#2#3{           {\it Ann. Phys. (NY) }{\bf #1} (19#2) #3}
\def\arnps#1#2#3{        {\it Ann. Rev. Nucl. Part. Sci. }{\bf #1} (19#2) #3}
\def\cnpp#1#2#3{        {\it Comm. Nucl. Part. Phys. }{\bf #1} (19#2) #3}
\def\apj#1#2#3{          {\it Astrophys. J. }{\bf #1} (19#2) #3}
\def\app#1#2#3{          {\it Astropart. Phys. }{\bf #1} (19#2) #3}
\def\asr#1#2#3{          {\it Astrophys. Space Rev. }{\bf #1} (19#2) #3}
\def\ass#1#2#3{          {\it Astrophys. Space Sci. }{\bf #1} (19#2) #3}
\def\aa#1#2#3{          {\it Astron. \& Astrophys.  }{\bf #1} (19#2) #3}
\def\apjl#1#2#3{         {\it Astrophys. J. Lett. }{\bf #1} (19#2) #3}
\def\ap#1#2#3{         {\it Astropart. Phys. }{\bf #1} (19#2) #3}
\def\ass#1#2#3{          {\it Astrophys. Space Sci. }{\bf #1} (19#2) #3}
\def\jel#1#2#3{         {\it Journal Europhys. Lett. }{\bf #1} (19#2) #3}
\def\ib#1#2#3{           {\it ibid. }{\bf #1} (19#2) #3}
\def\nat#1#2#3{          {\it Nature }{\bf #1} (19#2) #3}
\def\nps#1#2#3{        {\it Nucl. Phys. B (Proc. Suppl.) }{\bf #1} (19#2) #3} 
\def\np#1#2#3{           {\it Nucl. Phys. }{\bf #1} (19#2) #3}
\def\pl#1#2#3{           {\it Phys. Lett. }{\bf #1} (19#2) #3}
\def\pr#1#2#3{           {\it Phys. Rev. }{\bf #1} (19#2) #3}
\def\prep#1#2#3{         {\it Phys. Rep. }{\bf #1} (19#2) #3}
\def\prl#1#2#3{          {\it Phys. Rev. Lett. }{\bf #1} (19#2) #3}
\def\pw#1#2#3{          {\it Particle World }{\bf #1} (19#2) #3}
\def\ptp#1#2#3{          {\it Prog. Theor. Phys. }{\bf #1} (19#2) #3}
\def\jppnp#1#2#3{         {\it J. Prog. Part. Nucl. Phys. }{\bf #1} (19#2) #3}
\def\cpc#1#2#3{         {\it Comp. Phys. Commun. }{\bf #1} (19#2) #3}
\def\rpp#1#2#3{         {\it Rep. on Prog. in Phys. }{\bf #1} (19#2) #3}
\def\ptps#1#2#3{         {\it Prog. Theor. Phys. Suppl. }{\bf #1} (19#2) #3}
\def\rmp#1#2#3{          {\it Rev. Mod. Phys. }{\bf #1} (19#2) #3}
\def\zp#1#2#3{           {\it Zeit. fur Physik }{\bf #1} (19#2) #3}
\def\fp#1#2#3{           {\it Fortschr. Phys. }{\bf #1} (19#2) #3}
\def\Zp#1#2#3{           {\it Z. Physik }{\bf #1} (19#2) #3}
\def\Sci#1#2#3{          {\it Science }{\bf #1} (19#2) #3}
\def\n.c.#1#2#3{         {\it Nuovo Cim. }{\bf #1} (19#2) #3}
\def\r.n.c.#1#2#3{       {\it Riv. del Nuovo Cim. }{\bf #1} (19#2) #3}
\def\sjnp#1#2#3{         {\it Sov. J. Nucl. Phys. }{\bf #1} (19#2) #3}
\def\yf#1#2#3{           {\it Yad. Fiz. }{\bf #1} (19#2) #3}
\def\zetf#1#2#3{         {\it Z. Eksp. Teor. Fiz. }{\bf #1} (19#2) #3}
\def\zetfpr#1#2#3{    {\it Z. Eksp. Teor. Fiz. Pisma. Red. }{\bf #1} (19#2) #3}
\def\jetp#1#2#3{         {\it JETP }{\bf #1} (19#2) #3}
\def\mpl#1#2#3{          {\it Mod. Phys. Lett. }{\bf #1} (19#2) #3}
\def\ufn#1#2#3{          {\it Usp. Fiz. Naut. }{\bf #1} (19#2) #3}
\def\sp#1#2#3{           {\it Sov. Phys.-Usp.}{\bf #1} (19#2) #3}
\def\ppnp#1#2#3{           {\it Prog. Part. Nucl. Phys. }{\bf #1} (19#2) #3}
\def\cnpp#1#2#3{           {\it Comm. Nucl. Part. Phys. }{\bf #1} (19#2) #3}
\def\ijmp#1#2#3{           {\it Int. J. Mod. Phys. }{\bf #1} (19#2) #3}
\def\ic#1#2#3{           {\it Investigaci\'on y Ciencia }{\bf #1} (19#2) #3}
\def\tp{these proceedings}
\def\pc{private communication}
\def\opc{\hbox{{\sl op. cit.} }}
\def\ip{in preparation}
\thispagestyle{empty}
\begin{titlepage}
\begin{center}
\hfill hep-ph/9701420\\
\hfill FTUV/97-04\\
\hfill IFIC/97-04\\
\hfill IC/97/7\\
\hfill Junuary 30, 1997\\
\vskip 0.3cm
\large
{\bf Neutrino Conversions in a Polarized Medium}
\end{center}
\normalsize
\vskip1cm
\begin{center}
{\sl H. Nunokawa}$^1$ 
\footnote{
E-mail: nunokawa@flamenco.ific.uv.es},
{\sl V. B. Semikoz}$^1$ 
\footnote{E-mail: semikoz@izmiran.rssi.ru, 
On leave from the {\em Institute of the Terrestrial Magnetism, 
the Ionosphere and Radio Wave Propagation of the Russian Academy of 
Sciences, IZMIRAN, Troitsk, Moscow region, 142092, Russia}}
\vs .2cm
{\sl A. Yu. Smirnov}$^2$
\footnote{
E-mail: smirnov@ictp.trieste.it}, and
{\sl  J. W. F. Valle}$^1$
\footnote{E-mail: valle@flamenco.ific.uv.es}\\
\end{center}
\begin{center}
\baselineskip=13pt
$^1$ {\it Instituto de F\'{\i}sica Corpuscular - C.S.I.C.\\
Departament de F\'{\i}sica Te\`orica, Universitat de Val\`encia\\}
{\it 46100 Burjassot, Val\`encia, SPAIN         }\\
\vglue 0.8cm
$^2$ {\it International Centre for Theoretical Phsyics\\
P. O. Box, 586, 34100 Trieste, Italy\\}
\vglue 0.8cm

\end{center}
\vglue 1.0cm
\begin{abstract}

Electron polarization induced by magnetic fields can modify
the potentials relevant for describing neutrino conversions
in media with magnetic fields. The magnitudes of polarization
potentials are determined for different conditions.
We show that variations of the electron polarization along
the neutrino trajectory can induce resonant conversions in
the active-sterile neutrino system, but cannot lead to level
crossing in the active-active neutrino system. For neutrino
flavour conversions the polarisation leads only to a shift
of the standard MSW resonance.
For polarizations $\lambda \lsim 0.04$ the direct modifications 
of the potential (density) due to the magnetic field pressure 
are smaller than the modifications due to the polarization effect.
We estimate that indeed the typical magnitude of the polarization
in the sun or in a supernova are not expected to exceed $10^{-2}$.
However even such a small polarization may lead to interesting
consequences for supernova physics and for properties of neutrino
signals from collapsing stars.

\end{abstract}

\vfill

\end{titlepage}

\baselineskip = 0.5cm

\section{Introduction}
\vskip 0.3cm

Neutrino propagation in magnetized media has attracted considerable
attention recently, both from the point of view of the early universe
cosmology as well as astrophysics.
The presence of magnetic fields in the Universe \cite{Zeldo}
as well as in various astrophysical objects \cite{Raffelt0}
can affect neutrino conversion rates and this could have
important implications.

It has been shown that the neutrino dispersion relations in media with
non-zero magnetic fields are modified with respect to those in vacuo
and that the effect of the magnetic field can be equivalently described
as a correction to the neutrino self-energy \cite{disp,Raffelt}.
An alternative equivalent approach to this problem has been given
in ref. \cite{SemikozValle} where the matrix element of the axial
vector current has been calculated for an electron-positron plasma
in the presence of a magnetic field.
The basic result is that the potential relevant for neutrino
propagation acquires a component (axial potential) which is
proportional to the scalar product of the neutrino
momentum and the magnetic field vector.
In particular, it has been claimed that the axial potential
can be larger than vector current term, thus inducing the
possibility of a new type of resonant neutrino conversions
\cite{ssv,DOlivo}.

Several papers have considered the implications of the
axial potential in the propagation of neutrinos in media
which may be magnetized either by regular or random magnetic
fields. In the latter case neutrino conversions become aperiodic 
\cite{SemikozValle}.
The effect of such axial potentials on active-sterile supernova 
and early universe neutrino conversions has been discussed \cite{sergio}. 
In addition, the corresponding effect of strong random magnetic 
field upon neutrino transitions induced by a transition magnetic 
moment in the early universe hot plasma or in a supernova was
discussed in ref. \cite{sergio1}.
It has also been realized recently \cite{Kusenko} that even
small polarization effects may have very remarkable implications.
For example, they may lead to an explanation for the birth
velocities of pulsars as resulting from asymmetries due to
neutrino conversions in the cooling protoneutron star.

In this paper we consider some general features of the
neutrino propagation in polarized media. We generalize
the results of an early computation of the polarization
effect in ref. \cite{Langacker}. In particular, we show
that the previously considered magnetization effects can
be equivalently treated as the effect of {\it polarization} 
of the electrons induced by the magnetic field.

We find the potentials in terms of the averaged polarization
of the medium (sect. 2) and then calculate the averaged
polarizations in the magnetic field for various 
physical conditions (sect. 3.).  This approach gives a more
transparent physical interpretation of the results and allows
one to obtain an important restriction which was missed in ref.  
\cite{ssv,DOlivo}, a fact which led to some incorrect statements.
In sect. 4 we study the influence of the polarization on
neutrino conversions.  We show that electron polarization
can induce resonant conversions in the active-sterile neutrino
system, but can lead only to a shift of the usual MSW resonance 
\cite{MS,Wolfenstein} for neutrino flavour conversions.
In sect.5 we consider possible implications of the polarization
effects for solar and supernova neutrinos. We have explored
quantitatively the expected magnitude of the polarization
which is consistent with realistic density, $Y_e$ and
temperature profiles found in the sun or a supernova.  
We have estimated that for polarizations $\lambda \lsim 0.04$ 
the  modifications in the potential are smaller than the corresponding 
direct modifications of the potential due to field pressure.
We estimate the typical magnitudes of the polarization for
various physical situations.
Even though we find that the expected values of the
polarization is small, it can lead to interesting
consequences for supernova physics and for properties
of neutrino signals from collapsing stars.

\section{Effective potentials in a polarized medium}
\vskip 0.3cm

In what follows we will consider mainly polarization effects of 
electrons in a medium. In most cases the polarization of nuclei 
and nucleons is much weaker. Moreover, in the standard model only
electrons 
are relevant for conversions involving only active neutrinos. 
We will consider nucleons polarization in section 3.3 and 3.5. 

The effect of the medium is described by the potential  
\beq
V  = \langle  \Psi | H_{int} |\Psi \rangle ~, 
\eeq 
where  $\Psi$ is the wave function of the system neutrino-medium, and 
$H$ is the  standard Hamiltonian of the weak 
interaction at low energies
\beq
H_{int} = \frac{G_F}{\sqrt{2}} \nu \gamma^{\mu}(1 - \gamma_5)\nu 
\bar{e}\gamma_{\mu}(g_V + g_A \gamma_5) e ~.
\eeq
Here $G_F$ is the Fermi 
constant and $g_V$ and $g_A$ are the effective vector and 
axial-vector  
coupling constants in the standard model.

Let us consider the propagation of ultra-relativistic neutrinos with 
helicity - 1 in  medium with free electrons having the distribution 
(density)  
$$
\frac{f(\vec{\lambda}_e,\vec{p}_e)}{(2\pi)^3} 
$$  
over the momentum,  $\vec{p}_e$, and the polarization $\vec{\lambda}_e$. 
The vector of polarization is determined as   
\beq
\vec{\lambda}_e =   \omega_e^{\dagger} \vec{\sigma}\omega_e ,  
\eeq 
and $\omega_e$ denotes the two-component spinor  of the electrons. 
The total number density  of electrons, $n_e$, equals 
\beq 
\label{ne}
n_e = \sum_{\vec{\lambda}} 
\int  \frac{d^3 p_e}{(2\pi)^3} ~f(\vec{\lambda_e},\vec{p}_e).
\eeq

In an {\it unpolarized} medium, $\vec{\lambda}_e = 0$, the potential 
is determined by the vector component of the electron current: 
\beq \label{potvec}
V = V^V(\vec{p}_e)  = \sqrt{2} G_F\, g_V  \frac{f_e (\vec{p}_e)}{(2\pi)^3}  
\left(1  - \frac{\vec{p}_e \cdot \widehat{k}_{\nu}}{E_e} \right) , 
\eeq
where $\widehat{k}_{\nu} \equiv \vec{p}_{\nu}/|\vec{p}_{\nu}|$ with 
$\vec{p}_{\nu}$ being the neutrino momentum, $E_e$ is the energy of 
electrons. The expression in \eq{potvec} should be integrated over 
the $\vec{p}_e$ distribution of electrons. In an isotropic medium the 
second term is averaged out and we get the usual formula \cite{Wolfenstein}
\beq \label{pottotal1}
V  = \sqrt{2} G_F\,g_V n_e ~,   
\eeq
where the total concentration $n_e$ is determined in \eq{ne}. 
\Eq{pottotal1} also holds for the anisotropic case when the 
fluxes of electrons moving in opposite directions are equal. 
Such a situation is realized in a magnetized medium (see sect.3).   

In the case of a {\it polarized} medium the axial vector current also 
contributes \cite{Langacker}. Performing a straightforward calculation
we get the general expression   
\beq 
\label{potaxial}
V^A(\vec{\lambda}_e, \vec{p}_e)  = \sqrt{2} G_F\, g_A  
\frac{f(\vec{\lambda}_e,\vec{p}_e)}{(2\pi)^3} 
\left[ 
\frac{(\vec{p}_e \cdot \vec{\lambda}_e )}{E_e} -  
\frac{m_e}{E_e} (\widehat{k}_{\nu} \cdot \vec{\lambda}_e) - 
\frac{(\vec{p}_e \cdot \vec{\lambda}_e)(\vec{p}_e \cdot \widehat{k}_{\nu})}
{E_e (E_e + m_e)} 
\right] , 
\eeq
Note that $V^A \propto \lambda_e$,  and therefore in an unpolarized 
medium $V^A = 0$.   

Let us consider some special cases of \eq{potaxial}. 

\begin{itemize}

\item 
For a {\it non-relativistic electron medium}, $\vec{p}_e \approx 0$, 
we get from  \eq{potaxial}
\beq 
\label{potaxnonr}
V^A  \approx - \sqrt{2} G_F\, g_A  n_e  
\biggl(\widehat{k}_{\nu} \cdot \vev{\vec{\lambda}_e}\biggr)~, 
\eeq
where the average polarization of electrons is defined as  
\beq \label{polave}
\vev{\vec{\lambda}_{e}} = 
\frac{1}{n_e}\ 
\sum_{\vec{\lambda}}
\int  \frac{d^3 p_e}
{(2\pi)^3} \vec{\lambda} ~
f(\vec{\lambda}_e,\vec{p}_e),
\eeq
The total potential is 
\beq 
\label{pottot}
V  =  \sqrt{2} G_F\, n_e  
[g_V -  g_A \biggl(\widehat{k}_{\nu}\cdot \vev{\vec{\lambda}_e}\biggr)] .  
\eeq

\item 

In the case of {\it ultra-relativistic electrons} we get from 
\eq{potaxial}
\beq 
\label{potaxur}
V^A  \approx \sqrt{2} G_F\, g_A  
\frac{f(\vec{\lambda}_e,\vec{p}_e)}{(2\pi)^3} 
(\widehat{k}_e \cdot \vec{\lambda}_e)  
\left[1 - (\widehat{k}_e \cdot \widehat{k}_{\nu})  
\right] ~.  
\eeq
Here $\widehat{k}_{e} \equiv \vec{p}_{e}/|\vec{p}_{e}|$ is the 
unit vector in the direction of electrons momentum. 
If electrons are polarized in the transverse plane the potential 
is zero for any momenta of neutrinos. 
The potential is suppressed if neutrinos and electrons 
are moving in the same direction. 

\item 
The expression for the potential is simplified if 
electrons have a certain helicity:  
\beq 
\label{potaxial2}
V^A(\vec{\lambda}_e = \pm \widehat{k}_e) = 
 \pm \sqrt{2} G_F\, g_A  
\frac{f(\vec{\lambda}_e,\vec{p}_e)}{(2\pi)^3} 
\left[ \frac{p_e}{E_e} -  
(\widehat{k}_e \cdot \widehat{k}_{\nu}) \right] .   
\eeq
In the ultra-relativistic case this expression 
is reduced to \eq{potaxur}  and in non-relativistic case we get 
\eq{potaxnonr} .  

\item 

The case which is important for a {\it magnetized medium}  
(see sect. 3) is when there are two equal electron fluxes moving 
in opposite directions but heaving the same polarization 
along the momentum (electrons in the lowest Landau level). 
Using \eq{potaxial2}  we find   
\beq 
\label{potaxlan}
V^A  = - \sqrt{2} G_F\, g_A  n_e  
(\widehat{k}_{\nu} \cdot \vec{\lambda}_e) .  
\eeq
Here $n_e$ is the total concentration of electrons in both 
fluxes. Let us underline that this relativistic expression 
coincides with the non-relativistic formula \eq{potaxnonr}.   

\end{itemize}

In what follows we will  consider the axial vector potential in 
\eq{potaxnonr} or \eq{potaxlan}. 

The total effective potential resulting from electrons, 
protons and neutrons in an electrically neutral medium can be 
written in the form, 
\beq 
\label{pottotal}
V  = \sqrt{2} G_F\, n_e \left[ g_V - g_A \widehat{k}_{\nu} \cdot 
\vev{\vec{\lambda}_e\,} \right] + \sqrt{2} G_F\, n_n g_V^n, 
\eeq
In \eq{pottotal}, the second term describes neutrino-nucleon
scattering, with $n_n$ being the neutron concentration.

The effect of the medium on neutrino propagation is determined by the 
difference of potentials. For the case of $\nu_e \ra \nu_\mu, \nu_\tau$ 
flavour conversion only charge current neutrino-electron scattering
gives a net contribution. Using $g_V = - g_A = 1$ one finds 
\begin{eqnarray}
\label{potflavor}
V_{e\mu}  & = &\sqrt{2} G_F\, n_e \left[ 1 +  \widehat{k}_{\nu} \cdot 
\vev{\vec{\lambda}_e\,} \right] \\ 
& =  &\sqrt{2} G_F\, n_e \left[ 1 +  \vev{\lambda_e}\cos \alpha 
\right]\nonumber, 
\end{eqnarray}
where $\alpha$ is the angle between the neutrino momentum and 
the average polarization of electrons. There is no effect of 
nucleons in this case. Depending on the direction of polarizations 
the axial term can either enhance or suppress the potential. 
The maximal effect is obtained in the case of complete polarization
in the direction of  the neutrino momentum, $\vev{\lambda} = 1$.
In the case of complete polarization against the 
neutrino momentum, $\cos \alpha = -1$, $\vev{\lambda} = 1$, 
we get $V_{e\mu}=0$. 
Clearly, the axial vector term can not overcome the vector term, 
$| V_V| \geq |V_A| $. Thus it can not change the sign of $V_{e\mu}$ 
and therefore it can not induce resonant conversions. 

Let us now assume also that some amount of positrons is present in 
medium. Then the axial part of the potential for  
$g_A = -1$ is given by,
\beq 
\label{potax}
V^{A}=\sqrt{2} G_F\,\widehat{k}_\nu\cdot
\left[ n_e\vev{\vec{\lambda}_{e}}+n_e^+\vev{\vec{\lambda}_{e^+}} 
\right],
\eeq
where $n_e^+$ is  the positron concentrations. 
Note that electron and positron contributions have the same sign, 
so that the net effect is determined  by the relative 
polarization of electrons and positrons. 
Due to the different signs of their electric charges 
electrons and positrons are polarized in opposite directions 
in the magnetic field and therefore,  
$\vev{\vec{\lambda}_{e^+}} =-\vev{\vec{\lambda}_{e}}$, so that 
\beq 
\label{potax2}
V^{A}=\sqrt{2} G_F\,\widehat{k}_\nu\cdot \vev{\vec{\lambda}_{e}}
\left[ n_e -n_e^+ \right].
\eeq
Note that the vector part also depends on the difference of electrons 
and positrons concentrations and thus the total potential is 
determined by $\Delta n \equiv  n_e - n_e^+ $. 

In the case of conversion into sterile neutrinos,
the difference of potentials has also the contribution
from neutrino-nucleon scattering. With unpolarized 
nucleons we find, 
\beq \label{potes}
V_{es}  = \sqrt{2} G_F\, n_e \left[ \left(1- \frac{n_n}{2n_e} \right) 
+ \frac{1}{2}\,\widehat{k}_{\nu} \cdot \vev{\vec{\lambda}_e\,} \right].
\eeq
Now the polarization term can be bigger than the vector one,
thus leading to the possibility of level crossing induced by the
axial term.  

In next section we will calculate $\vev{\vec{\lambda}_{e}\,}$ for 
different physical conditions.

\section{Polarization in a medium with magnetic field}
\vskip 0.3cm

Let us calculate the polarization $\vev{\vec{\lambda}_e \,}$ in a
medium with electrons and positrons in the presence of a magnetic field. 
Suppose that the magnetic field is in the positive $z$ direction, 
$\vec{B}=(0,0,B)$. In this case  the energy spectrum of electrons 
is quantized according to
\begin{equation}
\varepsilon (p_z,n,\lambda_z)=
\sqrt{p_z^2 + m_e^2 + eB(2n + 1 +\lambda_z)} \ \ 
(n=0,1,2,...., \  \lambda_z = \pm 1),
\label{spectrum}
\end{equation} 
where $p_z$ is the neutrino momentum in the $z$ direction,
$m_e$ is the electron mass, $e\,(>0)$ is the absolute value of the   
electric charge, and $\lambda_z$ is the $z$ component of
$\vec{\lambda}_e$.
One can rewrite \eq{polave} as
\beq 
\label{spin}
\vev{\lambda_{e}} = \frac{1}{n_e}
\sum^{\infty }_{n =0} \sum_{\lambda_z}
\frac{eB}{(2\pi )^2} 
\int^{\infty }_{-\infty} dp_z ~ \lambda_z f(p_z,n, \lambda_z),
\eeq
with 
\beq 
\label{ne2}
n_e = 
\sum^{\infty }_{n =0} \sum_{\lambda_z}
\frac{eB}{(2\pi )^2} 
\int^{\infty }_{-\infty} dp_z ~ f(p_z,n, \lambda_z).
\eeq
Here, we have replaced the integration over $dp_x dp_y$ by 
a summation over $n$, taking into account that 
$\int dp_x dp_y \ra \sum_n 2\pi eB$. 
We take for the distribution function $f(p_z,n, \lambda_z)$ 
as the usual Fermi-Dirac form
\beq 
\label{fd}
f(p_z,n,\lambda) = \frac{1}{\exp[ (\varepsilon (p_z,n,\lambda_z) 
-\mu)/T] + 1},
\eeq
where $\mu$ is the chemical potential of electrons and 
$T$ is the temperature. From \eq{spectrum} one sees 
that the energy spectrum of 
electrons consists of the lowest Landau level, $n=0, \lambda_z = -1$, 
plus pairs of degenerate levels with opposite polarizations.
As a result only the lowest level survives in the sum in \eq{spin} 
\cite{SemikozValle,Raffelt}, so that the average 
polarization of electrons can be expressed as, 
\beq \label{spin2}
\vev{\lambda_{e}} =  - \frac{n_0}{n_e} \equiv
\frac{1}{n_e} \frac{e B}{(2\pi )^2} 
\int^{\infty }_{-\infty} dp_z ~ 
\frac{-1}{\exp[  (\sqrt{p_z^2 + m_e^2}-\mu)/T] + 1}~ ,
\eeq
where $ n_0$ is the electron number density in the lowest 
Landau level. Similarly, for positrons, we obtain
\beq 
\label{spin3}
\vev{\lambda_{e^+}} = \frac{n_0^+}{n_e^+}
=\frac{n_0 (\mu \rightarrow -\mu)}{n_e (\mu \rightarrow -\mu)} ,
\eeq
where $n_0^+$ is the positron concentration in the 
lowest Landau level. From \eq{potax} the axial-vector potential 
induced by the polarization can be written (for $g_A = -1$) as, 
\beq 
\label{potax3}
V^{A} = - \sqrt{2} G_F\, ( n_0  - n_0^+ )\cos\alpha_B ~,
\eeq
where $\alpha_B$ is the angle between the neutrino momentum and the
direction of the magnetic field. Note that since the  
electrons in the lowest Landau level are polarized against the 
field, $\cos\alpha = - \cos\alpha_B$.  
Substituting the expression for $n_0^+$ and $n_0$ given 
in \eq{spin2} into \eq{potax3} we reproduce the same formula
for $V^A$ found by other methods in ref. \cite{SemikozValle} 
and in ref. \cite{Raffelt}
\footnote{Note that in ref. \cite{Raffelt} the result in
\eq{potax3} was obtained through the calculation of one-loop 
diagrams using electron Green functions  in the medium
in the presence of magnetic field.}. 

Let us show that the same expression for the potential 
\eq{potax3}  is true even for the general case of \eq{potaxial}.
Let us first calculate the contribution to potential $V^A$ from 
the electrons in the level characterized by $p_z, n, \lambda$. 
For this we should average general expression for 
$V^A(\vec{p}_e, \vec{\lambda}_e$) \eq{potaxial} , 
over the azimuthal angle of the electrons (this angle fixes 
direction of the electron momentum in the plane orthogonal to 
the magnetic field).  Averaging over two possible 
directions of momentum $\pm p_z$ we find 
\beq 
\label{potbar}
\bar{V}^A (p_z, n, \lambda_z) =  
\frac{1}{4 \pi}
\sum_{\pm p_z} \int  d \phi V^A (\vec{p_e}, \vec{\lambda}_e) =  
- \sqrt{2} G_F\, g_A  \lambda_z \cos \alpha_B 
\left[1 -  
\frac{E_e^2 - p_z^2 - m_e^2}{E_e (E_e + m_e)}  
\right]~,   
\eeq
where $E_e = \varepsilon (p_z, n, \lambda_e)$ is given in  
\eq{spectrum} . 
Summing over all the levels we get total axial vector 
potential: 
\beq \label{potax5}
V^A  = 
\sum^{\infty }_{n =0} \sum_{\lambda_z}
\frac{eB}{(2\pi )^2} 
\int^{\infty }_{-\infty} dp_z ~ 
\bar{V}^A (p_z,n, \lambda_z) ~f(p_z,n, \lambda_z) .
\eeq
The contributions from the levels with the same energy 
and opposite $\lambda_z$ cancel each other, and  
the effect is determined by the lowest Landau level. 
For this level $\lambda_z = -1$ and 
$E_e^2 = p_z^2 + m_e^2$,  so that second term in \eq{potbar} 
is zero and integration becomes trivial:  
\beq 
\label{potfin}
V^A  =  - \sqrt{2} G_F\, g_A n_0 \lambda_e \cos \alpha_B  
\eeq
which coincides with \eq{potax3}.

One remark is in order. The expressions for the potentials 
in terms of averaged polarizations have been obtained for 
free-electron wave functions. These expressions can also be used
in the weak magnetic field limit: $eB \ll \vev{p^2_z}$. In general, 
for strong magnetic fields the use of free-electron wave functions 
is not justified (if fact, in  \eq{potbar} and  \eq{potax5} we have
used the modified dispersion relation \eq{spectrum}). In our case, 
however, the task is simplified since the polarization potential
is determined only by electrons at the lowest Landau level. In this 
level the electrons are moving along the magnetic field, they obey
the vacuum relation between the momentum and energy, and are 
described by free-electron wave functions. Therefore we can 
apply immediately the results of sect. 2 even for the case
of strong magnetic fields.

\subsection{Strongly degenerate electron gas}
\vskip 0.3cm

Let us now determine the expression for the electron concentration 
at the lowest Landau level $n_0$ for the case of a strongly degenerate 
electron gas: $(\mu - m_e)/T \gg 1$. 
In this case the Fermi-Dirac distribution \eq{fd} 
can be approximated by the  step function: 
\begin{equation}
\label{step}
[\exp ((\sqrt{p_z^2 + m_e^2 + 2eBn} - \mu)/T) +
1]^{-1}\to \theta (\mu  - \sqrt{p_z^2 + m_e^2 + 
2eBn})~  
\end{equation} 
(for the levels with $\lambda = - 1$; due to degeneracy 
it is enough to consider the levels with this polarization).  
Using approximation \eq{step} we obtain from \eq{spin2} 
\beq 
\label{nzero}
n_0 = \frac{eB p_F}{2\pi^2},
\eeq
where 
$p_F = \sqrt{\mu^2  - m^2_e}$. The Fermi momentum  $p_F$ 
is determined from the expression for 
the total electron concentration $n_e$ which can be 
obtained by the explicit integration in \eq{ne2} 
with \eq{step}:   
\begin{equation}
n_e = 
\frac{eBp_{F}}{2\pi^2} 
+ \sum_{n =
1}^{n_{max}}\frac{2eB\sqrt{p_{F}^2 - 2eBn}}{2\pi^2} 
\label{netotal}.
\end{equation}
The factor 2 in the sum takes into account the degeneracy of 
levels. Note that in general $p_F$ depends on $B$. 
In \eq{netotal} the first term is the contribution from the lowest 
Landau level and the second one results from the summation over 
all higher Landau levels. The summation goes up to a maximum 
value $n_{max} = [ p_F^2/(2eB) ]$, the integer part of $p_F^2/(2eB)$. 
If the magnetic field is very strong, $2eB \geq p_{F_e}^2$, 
the sum  vanishes and all electrons are at the main Landau level, 
which means that the electron gas is fully polarised and all the 
electron spins are aligned opposite to the magnetic field. 
From \eq{nzero} and \eq{netotal} we find 
\begin{equation}
\label{nzero2}
n_0 = \frac{n_e}
{
1+ \sum_{n = 1}^{n_{max}}  
\sqrt{1  - \frac{2eBn}{ p_{F}^2}} 
}.
\end{equation}
Clearly $n_e \geq n_0$.
\vglue 0.3cm
Let us now consider two extreme cases. 
\vglue 0.3cm
\noindent
\ben
\item
 Strong magnetic field limit:
\beq 
\label{condition0}
2eB > p_F^2.
\eeq
The sum in \eq{nzero2} disappears so that $n_e = n_0$, i.e. all the 
electrons are in the first Landau level: the medium is completely polarized. 
In this case we get from \eq{netotal} 
\beq 
\label{fermi1}
p_F= \frac{2\pi ^2 n_e}{eB}~, 
\eeq
and the condition \eq{condition0} of complete polarization 
can be written as  
\beq 
\label{condition00}
B > \frac{1}{e}\left(\sqrt2 \pi^2 n_e \right)^{2/3}~. 
\eeq
Clearly, the higher  the density the larger magnetic field 
required in order to have complete polarization.
\vglue 0.4cm
\noindent
\item
 Weak field limit: 
\beq 
\label{weaklimit}
eB \ll p_F^2.
\eeq
The the sum in \eq{netotal} contains contributions from 
many Landau levels and dominates over the first term. 
The sum can be approximated by integration as follows, 
\beq \label{weak}
n_e = \frac{eBp_{F}}{2\pi^2}\left[ 1 +
\frac{p_{F}^2}{eB}\int^1_{2eB/p_{F}^2}dx\sqrt{1 - x}\ \right]
\approx \frac{p_{F}^3}{3\pi^2}
\Bigl [ 1  + \frac {3eB}{2p_{F}^2}\Bigr ]~.
\eeq
From this  we get the usual expression for $p_F$  in a medium 
without magnetic field: 
\beq 
\label{fermi2}
p_F \simeq (3\pi^2 n_e)^{1/3}~. 
\eeq
Inserting this $p_F$ in \eq{nzero} we have, 
\beq 
\label{nz}
n_0 =  \frac{eB}{2}\left(\frac{3n_e}{\pi^4}\right)^{1/3} ~,
\eeq
and consequently, 
\beq 
\label{pol2}
\vev{\vec{\lambda}_e} =
- \frac{e\vec{B}}{2} \left(\frac{3}{\pi^4}\right)^{1/3} n_e^{-2/3}.
\eeq
The polarization effect increases linearly with $B$ 
and decreases as $n_e^{-2/3}$. Using \eq{pol2} and \eq{potax3} 
we get for the effective potential of the electrons,
\beq 
\label{potax4}
V = \sqrt{2}G_F\,n_e -  
\frac{G_F\, eB}{\sqrt{2}} \left(\frac{3n_e}{\pi^4}\right)^{1/3} 
\cos\alpha_B.
\eeq
This expression coincides with one used in \cite{DOlivo,Kusenko}.  
Although one might think from \eq{potax4} \cite{ssv,DOlivo} that 
the axial contribution may be dominant, it should be clear from
our discussion above that  this expression is correct only if 
the polarization term is small in comparison with vector 
current contribution. 
\een
The general behaviour of the polarization $\vev{\lambda_e}=n_0/n_e$ with
respect to the magnetic field and density can be obtained 
from \eq{nzero2} for the case of strong degeneracy. The
result is shown in Fig. 1. In this figure the lines of 
equal polarization basically correspond to the dependence 
given in \eq{pol2}
\footnote{The small jumps in these lines for $\vev{\lambda_e}$ arise
 from discrete changes in the sum \eq{nzero2}.}.

\subsection{Electrons at finite temperature}
\vskip 0.3cm

The polarization effect decreases as the temperature  
of the medium increases. 

\bi
\item
For small finite temperatures,  $T \ll \mu -m_e $, 
 temperature effects give only a small negative contribution 
to the electron density in the first Landau level \cite{Landau}, 
\beq 
\label{temp}
n_0 \sim \frac{eB p_F}{2\pi^2}\Bigl [ 1 -
\frac{\pi^2}{6}\Bigl(\frac{m_eT}{p_F^2}\Bigr )^2 +...\Bigr ]~.
\eeq 
\item
For the opposite case of weak degeneracy, $(\mu-m_e)/T < 1$,
strong polarization is achieved if the first term in \eq{ne2} 
dominates over the higher levels. In particular, 
\beq 
\label{condition1}
\frac{1}{\exp[(m_e-\mu)/T] +1 } \gg
\frac{1}{\exp[(\sqrt{m_e^2 +2eB}-\mu)/T] + 1}~. 
\eeq
From this we get, 
\beq 
\label{condition2}
\frac{\sqrt{m_e^2 +2eB}-\mu}{T}\gg 1.
\eeq
If this condition is satisfied, the contribution to the total 
density from the higher Landau levels ($n=1,2,3...)$ is strongly 
suppressed for all momenta $p_z$, and as a result nearly
complete polarization is achieved, $n_e\sim n_0$.
Again here we identify two cases:
\ben
\item
In the weak field limit, $m_e^2 \gg 2eB$, the condition 
in \eq{condition2} becomes,
\beq 
\label{conditionweak}
\frac{eB}{m_eT} = \frac{2\mu_BB}{T}\gg 1,
\eeq
where $\mu_B$ is the Bohr magneton. According to \eq{conditionweak} 
the  interaction energy with the magnetic field (Zeeman energy) should 
be much bigger than the kinetic energy of the electrons. 
Polarization itself can be estimated as, in the non-relativistic case, 
\beq
\label{polel}
\vev{{\lambda}_e}  \sim \frac{\mu_{B} B}{T}~.  
\eeq
In ultra-relativistic case:
\beq
\label{polelrel}
\vev{\lambda_{e}} \sim 
\frac{e  B}{6T^2} =   
\frac{\mu_{B} B m_e }{3T^2}~.  
\eeq
In non-relativistic case the total potential becomes:
\beq
\label{potnonrel}
V \approx \sqrt 2 G_F n_e \left[ 1 - \frac{\mu_B B}{T} \cos \alpha_B
\right]
\eeq

\noindent
\item
In the strong field limit, $m_e^2 \ll 2eB$, we get from 
\eq{condition2},  
\beq 
\label{newparam2}
\frac{\sqrt{2eB}}{T} \gg 1.
\eeq
In these cases, there is a strong polarization even if 
electrons are not degenerate. 
\een
\ei

The general temperature dependence of the polarization $\vev{\lambda_e}$ 
for different values of the magnetic field is shown in Fig. 2a and 2b.
One sees that the depolarization effect due to temperature becomes 
strong for small values of the degeneracy parameter $(\mu-m_e)/T$. 
Note that even for very strong magnetic field, e.g. $2eB/p_F^2=2$,  
the polarization is only $\sim$ 5\% for $T \sim$ 1 MeV and 
$\rho Y_e = 10^6$ g/cm$^3$. For higher density, e.g.
$\rho Y_e = 10^8$ g/cm$^3$ and fixed value of  $2eB/p_F^2=2$, 
the depolarization effect is small for $T < $ 1 MeV.

\subsection{Polarized nucleons} 

For system of active-sterile neutrinos also the polarization of 
nucleons gives some effect. Since for sterile neutrinos $V^V = V^A = 0$,  
the difference of the potentials is determined by the total potential of 
the active component. 

Let us consider an electrically neutral medium which consists of electrons 
and non-relativistic protons and neutrons with concentrations $n_p$ and $n_n$ 
correspondingly. 
The total axial vector potential for the electron neutrino can be written 
as 
\beq 
\label{potaxnuce}
V^A  \approx - 
\frac{G_F}{\sqrt{2}} 
\widehat{k}_{\nu} \cdot \left[
- n_e \biggl(\vev{\vec{\lambda}_e} +  g_A^N \vev{\vec{\lambda}_p}\biggr) 
+ g_A^N n_n \vev{\vec{\lambda}_n} 
\right]~, 
\eeq
where $g_A^N \approx 1.26$ is the renormalization constant of the 
axial-vector nucleon current.  For the muon (or tau) neutrinos 
the electron contribution changes, and 
we get  
\beq 
\label{potaxnucmu}
V^A  \approx - 
\frac{G_F}{\sqrt{2}} 
\widehat{k}_{\nu} \cdot \left[
n_e \biggl(\vev{\vec{\lambda}_e}  - g_A^N  \vev{\vec{\lambda}_p}\biggr) 
+ g_A^N n_n \vev{\vec{\lambda}_n} 
\right]~, 
\eeq
where  
$\vev{\vec{\lambda}_p}$ and $\vev{\vec{\lambda}_n}$ 
are  the averaged polarizations of protons and neutrons. 

Note that according to \eq{nzero} and \eq{pol2} in the limit of a 
strongly degenerate distribution, the polarization does not depend 
on the mass of the particle and one would expect the same degree of 
polarization both for electrons and protons: 
$\vec{\lambda}_p \approx - \vec{\lambda}_e$.
However, in realistic conditions nucleons are typically strongly 
non-degenerate and their polarization can be estimated as 
\beq
\label{polnuc}
\vev{{\lambda}_{p, n}}  \sim \frac{\mu_{p,n} B}{T}~,  
\eeq
where  the magnetic moments of proton and neutron equal  
$\mu_p = 2.79 \mu_N$ and 
$\mu_n = -1.91 \mu_N$ 
with $\mu_N \equiv e/2 m_N$ 
(here $m_N$ is the mass of nucleon).  

In the magnetic field due to different signs of the magnetic moments 
protons and neutrons are polarized in different directions. 
Therefore according to \eq{potaxnuce}, \eq{potaxnucmu} the contributions 
of the protons and neutrons to the potentials have the same sign. 
Moreover, in muon neutrino potential all contributions sum.

If both electrons and neutrons are non-degenerate, then 
${\lambda}_N / {\lambda}_e \sim 10^{-3}$ and the polarization 
effect of the nucleons is about three orders of magnitude smaller. 
However, the degeneracy suppresses polarization so that if the 
electron distribution is strongly degenerate, whereas at the same
time the nucleons are strongly non-degenerate (as realized in central 
regions of hot neutron stars) the polarization of nucleons might
become important.

\section{Neutrino transitions in a polarized medium}
\vskip 0.3cm

The effect of the axial vector term in the resonant conversions 
of active \neus was first obtained in ref. \cite{ssv} and independently 
in \cite{DOlivo}. Here we will confine ourselves to the qualitative 
effects associated with polarization, and with a rough estimate of 
the magnitude of the polarization matter potentials for various
physical systems. The  explicit calculation of the corresponding 
conversion probabilities will be taken up elsewhere.

For definiteness, we consider a system of two neutrinos  
$\nu_e$ and $\nu_x$, where $\nu_x$ is an active neutrino 
($ \nu_\mu$ or $\nu_\tau$) or a sterile state $\nu_s$. 
The evolution equation is given by 
\beq 
\label{initial}
i\frac{d}{dt}\left (
\matrix{\nu_e\cr  \nu_x}\right )
 = \left ( \matrix{V-\Delta\cos2\theta & \frac{1}{2}\Delta\sin2\theta \cr
          \frac{1}{2}\Delta\sin2\theta & 0 }
       \right )\left (\matrix{\nu_e\cr\nu_x}\right ),   ~
\eeq
where $\Delta \equiv \frac{m_2^2 - m_1^2}{2E}$ with $m_2$ and $m_1$ 
being the neutrino masses, and $E$ is the neutrino 
energy\footnote{For the case of anti-neutrinos 
the evolution equation is the same but with opposite sign of $V$.}.

The effective potential $V$ for $\nu_e-\nu_{\mu,\tau}$ conversions given 
in previous sections can be written as,  
\beq 
\label{pot2}
V=V_{e;\mu ,\tau}=\sqrt{2} G_F\, n_e (r)[1 - b(\vec{r})\cos\alpha(\vec{r})],
\eeq
where
\beq
\label{polterm}
b (\vec{r}) \equiv \frac{n_0(\vec{r})  - n_0^+(\vec{r}) } {n_e(r)-n_e^+(r)}.  
\eeq
characterizes the  polarization term. For simplicity we have
assumed an isotropic electron density profile. Clearly the
axial term  changes the matter potential as, 
\beq
\label{potpol}
V \Bigl( n_e(r) \Bigr) \rightarrow 
V= V^V \Bigl( n_e(r) \Bigr) + 
V^A \Bigl( n_e(r), B(\vec{r}),
\alpha(\vec{r}) \Bigr)~. 
\eeq
Polarization can enhance or suppress the potential depending 
on the direction of the magnetic field. 
The modification of the potential is of the order 1 if the medium
is strongly polarized. If fact, one can envisage an extreme
situation in which in some region there is a complete 
polarization in the direction of  neutrino momentum.
In this case the polarization term cancels the vector term, 
leading to $V=0$. In such a region the evolution of the 
neutrino system is reduced to that in vacuo.

Let us now consider some possible  
effects of the polarization. 
\begin{itemize}
\item

Shift of the resonance.\\

The resonance condition, $V-\Delta \cos2\theta=0$ can be written as, 
\beq 
\label{res}
\sqrt{2} G_F\, n_e(r) [1 - b(r)\cos\alpha(r)] -\Delta \cos 2\theta =0. 
\eeq
There are two ways of interpreting the effects of the polarization.  
\ben
\item
Shift of the resonance position.  

It is clear that the presence of the polarization term changes the value 
of $r_R$ at which the resonance condition is fulfilled. In other words, 
the layer in which the flavour transition takes place is shifted depending 
on the strength and direction of the magnetic field. 
\item
 Shift of the neutrino parameters. 

According to \eq{res} the axial vector contribution changes the neutrino 
parameter ($\Delta$ - mass squared difference  or energy required for 
resonance in a certain layer with fixed total density $n_e$):  
$$
\Delta \ra \frac{\Delta}{(1-b \cos \alpha)}.
$$
Depending on the polarization direction the parameter $\Delta$ can be
diminished or increased.  
This shift may have a number of interesting consequences already  
discussed in ref. \cite{ssv,DOlivo}. For example,  a 
system of almost degenerate neutrinos can undergo resonant 
conversions in a strongly polarized medium in the case of 
positive (but  small) $\Delta$. Note that $\Delta$ can not be 
too small or zero since the mixing of neutrinos is proportional 
to $\Delta$ (see \eq{initial}) and with diminishing $\Delta$ 
adiabaticity starts to be broken. 

Note that for a system of active neutrinos the possibility of resonant 
conversion exists only for neutrinos or anti-neutrinos even if a medium 
is polarized: neutrinos and anti-neutrinos do not simultaneously convert, 
in contrast with the situation considered in ref. \cite{massless}. 
Moreover, since the polarization term can not overcome the vector potential 
term it can not induce resonant anti-neutrino flavour conversions if 
$\Delta > 0$, only for $\Delta < 0$.
This conclusion is in conflict with the papers in ref. \cite{ssv,DOlivo}.
Unfortunately, as we will see in the  next section, 
the level of 
polarization required in order to have strong  resonance shift  
 is too large to achieve with reasonable values of the 
magnetic field, at least for the case of the sun.
\een

\item 
Modification of the adiabaticity.\\ 

The polarization term  modifies the dependence of $V$ on $r$ and 
therefore, influences the adiabaticity of propagation. This in turn 
changes the transition probability in certain ranges of neutrino 
parameters $\theta$ and $\Delta m^2$. 

\item 
Noisy media.\\ 

The magnetic field may have a domain structure with 
different strength and direction in different domains. This leads 
to a perturbation of $V(r)$ profile which may have a random 
character. If the typical domain  size is smaller than the 
neutrino oscillation length, this modulation can be considered 
as density (potential) random fluctuations, similar to those
considered in ref. \cite{BalantekinLoreti,noise}.

\item 

Resonant conversion of sterile neutrinos driven by polarization 
effect.\\

Using the effective potential \eq{potes} for a system of 
active-sterile neutrinos we can write the resonance condition as, 
\beq 
\label{res2}
\sqrt{2} G_F\, n_e(r) 
\left[1-\frac{n_n(r)}{2n_e(r)} - \frac{1}{2} 
b(r)\cos\alpha(r) \right]
-\Delta \cos 2\theta =0,
\eeq
where $b(r)$ is defined in \eq{polterm}. 
For $n_e = n_p < n_n$ the polarization term 
can be bigger than the vector current contribution, 
\beq 
\label{sterile}
1-\frac{n_n(r)}{2n_e(r)} < \frac{1}{2} 
b(r)\cos\alpha(r).  
\eeq
When this condition is fulfilled, the resonant conversion can occur 
even if the $\Delta$ term is negligible in \eq{res2}. In fact the
resonant conversion can be driven by the polarization. The 
resonance condition \eq{res2} can be re-written as,
\beq 
\label{sterile1}
\frac{1}{2} b(r)\cos\alpha(r) =
1-\frac{n_n(r)}{2n_e(r)} \mp 
\frac{\Delta \cos2\theta}{\sqrt{2}G_F\, n_e(r)},
\eeq
where the minus sign is for neutrinos and the plus sign 
is for anti-neutrinos. The ratio $n_n(r)/2n_e(r)$ may have 
a rather weak dependence on $r$. Then the level crossing 
will be due mostly to changes of $b(r) \cos\alpha(r)$
on the way of neutrinos. That is, 
the change of the strength and the direction of the magnetic 
field along the neutrino motion will lead to resonant conversion. 
Note that if $\Delta$ is small enough, this condition is satisfied 
due to the polarization term both for neutrinos and anti-neutrinos. 
In other words, the polarization can induce resonant conversions in a 
medium with fixed chemical composition ($n_n/n_e \sim$ const) 
both in the neutrino and anti-neutrino channels. Unlike the
active-active neutrino conversion case discussed above,
here there can be simultaneous conversions of neutrinos 
and anti-neutrinos, similarly to the massless neutrino
conversion mechanism \cite{massless}. Strictly speaking,
we need a finite value of $\Delta$ in order to have mixing 
and adiabaticity, though it may be negligible in the
resonance condition, \eq{sterile1}. 
 
\end{itemize}

\section{Astrophysics}
\vskip 0.3cm

Although for realistic applications a detailed study
of the evolution equation \eq{initial} is needed, one can 
gain some insight by first performing simple estimates of 
the magnitude of the polarization for different systems.

\subsection{Polarization in the Sun}
\vskip 0.3cm

In the sun the degeneracy of electrons is small and the magnitude 
of the polarization is determined by the interplay of the magnetic 
field strength and the temperature:
$ 
\vev{\lambda_e} \sim \mu_B B / T
$ . 
Taking for the central region of the sun a maximal possible strength of 
the magnetic field, $B\sim 10^8$ Gauss, we get 
$\vev{\lambda_e} \sim 4\cdot 10^{-4}$. 
In the bottom of the convective zone $B$ could be as large as 
$10^6$ Gauss, leading to $\vev{\lambda_e} \sim 3 \cdot 10^{-5}$. 
Thus in the sun the magnitude of the polarization is too small to 
expect sizeable effects. It can give only very small perturbations 
of the density (potential) profile. 

\subsection{Polarization in supernovae}
\vskip 0.3cm

Let us now estimate the possible allowed magnitude of the
polarization  $\vev{\lambda}_e$ in supernovae. Fig. 3a and 
3b illustrate typical density and temperature profiles in 
a supernova with 20 $M_\odot$ progenitor for two different 
moments of time \cite{wilson}.

In Figs. 4a and 4b we have plotted required magnetic field 
profiles for producing a given polarization, 
$\vev{\lambda_e}$ = 0.01, 0.1 and 0.99 for the density and 
temperature profiles shown in 
Fig. 3a and 3b. The degeneracy parameter, $(\mu-m_e)/T$ is also 
plotted. Clearly in the earlier epoch ($t=0.15$ sec after the core 
bounce) the degeneracy condition: $(\mu-m_e)/T >1$ is satisfied 
in the region $r \lsim 100$ with $\rho Y_e \sim 10^9$ g/cm$^3$. 
The degeneracy parameter increases to 2 or larger at $r \lsim$ 50 km. 
In this region we can estimate the magnetic field needed in
order to have strong polarization using \eq{condition00}, 
\beq 
\label{mag}
B \sim 10^{17} (\rho_{12} Y_e)^{2/3}\ \  \mbox{Gauss},
\eeq
where $\rho_{12}$ is the matter density in units of 
$10^{12}\mbox{g/cm}^3$. At $r\simeq 100$ km, where 
 $\rho Y_e \sim 10^{9}$ g/cm$^3$, one should have 
$B \gsim 3\cdot 10^{16}$ Gauss for the polarization 
term to be close to one. For $r \simeq 400$ km
($\rho Y_e \sim 10^{8}$ g/cm$^3$) the required field is 
$\sim 3\cdot 10^{15}$ Gauss. 
For external regions ($r>1000$ km) the degeneracy is rather 
weak and the depolarization due to temperature becomes important. 

As can be seen from Fig. 4b, for the later epoch, strong degeneracy 
is realized in even more central regions: $r<12$ km, i.e. practically in 
the neutrinosphere. For $r > 15$ km the depolarization due to temperature 
dominates. In Fig. 4b we also show the magnetic field profiles
leading to $\vev{\lambda_e}$ = 0.01, 0.1, 0.99. 

Comparing these results with usually accepted large-scale
magnetic field profiles:
\beq
\label{bprofile}
B(r) \sim B_0 \left(\frac{r_c}{r}\right)^n,
~~\: B_0 = 10^{12}-10^{14} \ \mbox{Gauss},~~ n = 2,3,~~ r_c = 10~ 
\mbox{km}
\eeq
we conclude that the magnitude of the polarization term
does not exceed 1 \% level, and they drop below 0.1 \% in 
external regions. 

With the profile given in \eq{bprofile} one may conclude
that the effect of polarization will give only small modifications
of the potentials and their influence on possible neutrino 
conversion in this region is expected to be weak. 
Note that in stars of smaller mass the degeneracy parameter is 
stronger and temperatures are lower. Therefore, 
the polarization for realistic 
magnetic fields can be  bigger than 1 \%.

\subsection{Random Magnetic Fields}
\vskip 0.3cm

Besides  global large-scale magnetic fields, a star can also have 
magnetic fields in small domains of size $r_D$ ($r_D \ll r$, 
where $r$ is the distance from the center to layers with domains). 
The field in different domains can be randomly oriented. 
The existence of domains with $r_D \sim $ 1 km at 
$r\gsim 10$ km has been considered in ref. \cite{DunkanThomson} 
in the context of neutron star dynamics. 
The effect of the axial potential induced by such strong random 
magnetic fields on active-sterile supernova and early universe 
neutrino conversions has been discussed in ref. \cite{sergio,sergio1}. 
The strength of the random field inside the domains can be much larger 
than the strength of a global field, so that the polarization 
effect could be correspondingly bigger. There can be even stronger 
fields in filaments, like super-conducting needles.

As we have discussed in section 4, these domains or needles 
will cause a modulation of the density profile which will have 
a {\sl noisy} character. If the number of domains along the neutrino 
path is large enough, even small modulations  at the \% level 
can lead to large (of order 1) changes of the transition 
probabilities \cite{BalantekinLoreti,noise,noisesn}. 

Another effect is that the rotation of the star would induce a time 
dependence of the neutrino signal, since in different moments of time
neutrinos directed to the earth cross different magnetic field domains.

\subsection{Nucleon polarization}

Let us finally estimate possible polarization effects of nucleons
in a supernova. Everywhere outside the core the nucleon gas is strongly 
non-degenerate. In the neutrinosphere with density 
$10^{11}-10^{12}$ g/cm$^3$, $Y_e \sim 0.1$ and temperature 
$T \sim 7$ MeV, we get 
$E_F - m_p \approx  p_F^2/2m_p \lsim 5 \times 10 ^{-2} $ MeV 
and ($E_F - m_p)/T \lsim (1-5) \times 10^{-2}$,  
that is, the nucleons are non-degenerate. Thus \eq{polnuc} can 
be used and for $B = 10^{14}$ Gauss we find 
$\vev{{\lambda}_p} \sim  10^{-4}$, 
$\vev{{\lambda}_n}\sim 8 \cdot 10^{-5}$. Even though the electron 
gas is degenerate, however, the level of polarization is 
still higher: $\vev{{\lambda}_e} \sim 10^{-3}$. 
In the neutrino-sphere the ratio of the axial potentials is given by,
\beq
\label{ratio1}
|V_e^A| : |V_p^A| : |V_n^A|  \sim  1 : 0.2 : 1.
\eeq
In the core at densities $10^{14}$ g/cm$^3$, $Y_e\sim 0.3$ 
and temperatures $T \sim 20$ MeV, the degeneracy parameters 
for nucleons is: 
($E_F- m_p)/T \lsim 1$, that is, nucleons 
are only weakly degenerate and one can still use \eq{polnuc} 
for estimations.  For $B = 10^{14}$ Gauss we get   
$\vev{{\lambda}_p}  \sim 4 \cdot 10^{-5}$, 
$\vev{{\lambda}_n} \sim 3 \cdot 10^{-5}$ 
but the electron polarization is also strongly suppressed:  
$\vev{{\lambda}_e}  = 3 \cdot 10^{-5}$. We can also estimate 
the ratio of the axial potentials in the core, 
\beq
\label{ratio2}
|V_e^A| : |V_p^A| : |V_n^A|  \sim  1 : 2 : 3.
\eeq
Thus in the central regions of the core all components 
give comparable contribution to the axial-vector potential 
which can be important for conversion of the active neutrinos 
into sterile neutrinos.  

\subsection{Pressure versus polarization}
\vskip 0.3cm

As follows from our discussion strong polarization effects 
imply a strong magnetic field. Such a magnetic field also 
produces a pressure, 
\beq 
\label{pb}
P_B = \frac{B^2}{8\pi}.
\eeq
This pressure by itself modifies the density distribution and 
therefore, will influence neutrino conversion as well as the
dynamics of the star. Let us compare $P_B$ with the pressure 
of the degenerate electron gas 
\beq 
\label{pgas}
P_{Gas} = \frac{1}{12\pi^2} \mu^4.
\eeq
The effect on the density (potential) profile is determined by ratio, 
$P_B/P_{Gas}$ which can be written for relativistic case
($\mu \sim p_F$) as 
\beq 
\label{ratio}
\left(\frac{\Delta V}{V}\right) 
\sim \frac{P_B}{P_{Gas}} 
= \frac{3}{32\alpha}
\left(\frac{2eB}{p_F^2}\right) ^2,
\eeq
where $\alpha \equiv e^2/4\pi$. For magnetic fields which satisfy
the strong polarization condition \eq{condition0}  we get 
\beq 
\label{ratio1}
\frac{P_B}{P_{Gas}} 
> \frac{3}{32\alpha}
\sim 13. 
\eeq
Thus the pressure associated with the magnetic field dominates 
over the matter pressure. The direct impact of the magnetic field 
on $V$ and on the dynamics of the star will be stronger than its 
indirect influence through polarization. With 
diminishing  magnetic fields 
the direct effect of the field ($\sim B^2$) decreases faster than 
the polarization effect ($\sim B$). The polarization effect 
in the weak field limit (see \eq{pol2}) equals:
\beq 
\label{ratio2}
\left(\frac{\Delta V}{V}\right)_{pol} 
\sim \lambda 
= \frac{3}{4}
\left(\frac{2eB}{p_F^2}\right).  
\eeq
Using this expression one can estimate  the direct effect as
\beq 
\label{ratio3}
\left(\frac{\Delta V}{V}\right)_{dir} 
\sim \frac{1}{6\alpha} \lambda^2. 
\eeq
Comparing \eq{ratio2} and \eq{ratio3} we find that 
the polarization effect becomes larger than the direct 
effect: $(\Delta V/V)_{pol} > (\Delta V/V)_{dir}$ if  
\beq 
\label{poleffect}
\lambda < 6 \alpha \sim 4 \times 10^{-2}.
\eeq
In other words, this happens when the magnitude of the 
polarization term is small. Of course, the effects are
quite different. The magnetic pressure can only diminish 
the matter density and the potential, whereas the polarization 
can also enhance the potential and, for the case of active-sterile 
neutrinos, it can even change the sign. Moreover, the polarization 
term, depending on the direction of the magnetic field, leads to 
anisotropy of the potential. The 
pressure depends on the absolute value of the field. 
Of course in very strong magnetic fields one should take 
into account both effects simultaneously. 

\subsection{A comment on pulsar velocities}
\vskip 0.3cm

Even though the  polarization effects in a protoneutron 
star are expected to be small, at the $\lsim$ 1 \% level, 
as we have seen above, they may lead to observable
consequences. 

The most remarkable is  
an explanation for the birth velocities of pulsars \cite{Kusenko}.
The polarization of medium by dipole type magnetic field leads to
asymmetric 
shift of the resonance layer inside the star in one hemisphere with 
respect to the other. This in turn results in anisotropy 
of properties of emitted
neutrinos. This anisotropy is the reason of the pulsar's kicks.   

The required value of the \neu mass-squared parameter for the  
mechanism is $\gsim 10^4$ eV$^2$. This value is
larger than cosmologically allowed (unless neutrinos have new 
decay or annihilation channels into majorons ) and
lies, in particular, outside the preferred range where it
can play a role of hot dark matter, $\lsim 10 - 50$ eV$^2$. 
One suggestion to diminish $\Delta m^2$ is to assume a strong 
polarization effect. Indeed, almost complete polarization in 
the direction of neutrino propagation would strongly suppress
the effective potential for fixed total density, so that the 
resonance condition can be satisfied between neutrinospheres 
\ne and \nt 
for much smaller masses \cite{Kusenko2}.
Let us comment on this possibility. 

The dependence of polarization on the 
$\rho Y_e$ 
parameter for fixed $T = 7$  MeV
 is shown in Fig. 5. 
From this figure one finds that 90 \% polarization (which allows 
one to suppress the potential in certain directions in the hot neutron star) 
requires a field as strong 
as $B\gsim 10^{17}$ Gauss for $\rho Y_e \sim 10^{11}$ g/cm$^3$. 
As follows from sect. 5.4 for such a strong fields the magnetic 
field pressure will dominate and can strongly change the dynamics 
of collapse. 

Note that polarization does not depend on $\rho Y_e$ (for fixed T)  
at small densities and is determined by the magnetic field and 
temperature. The asymptotic value of polarization at small $\rho Y_e$  
are described by \eq{polelrel}.
\vskip 0.3cm
\section{Conclusions}
\vskip 0.3cm

\noindent
\ben
\item
We have shown that neutrino propagation in a magnetized medium 
\cite{disp,ssv,DOlivo,sergio,sergio1} can be equivalently seen
as being associated with the scattering of neutrinos on 
electrons polarized by the magnetic field. This  approach 
gives a more transparent physical interpretation of the effect 
and allows one to obtain important restrictions. 
\item
We have shown that, for neutrino flavour conversions the polarization 
term of the potential can not overcome usual vector current term. 
The polarization term can lead  to lowering or raising
(depending on the direction of polarization) of the the 
resonant density and, therefore, shift the position of the 
resonance layer. In contrast, in the case of active-sterile 
neutrino mixing, the polarization term can be bigger than the 
vector term. Thus the polarization can induce resonant neutrino 
conversions, even for small values of the parameter $\Delta$ in 
\eq{initial}. Moreover, in  media with fixed chemical 
composition the resonant conversions can take place both in 
neutrino and anti-neutrino channels. 
\item
 For realistic magnetic fields in the sun or in a supernova
the polarization does not exceed (0.1-1) \%. Although small,
the shift of the resonance layer can lead to an explanation 
of observed peculiar velocities of pulsars. 
\item
Strong magnetic fields may lead via pressure to a direct 
perturbation of density and therefore, the potential profile. 
The effect of such direct perturbation becomes stronger than 
the polarization effect we have studied for $\lambda \gsim 0.04$. 
\een

\newpage
\noindent {\bf Acknowledgment}\\
\vskip 0.3cm

We thank E. Akhmedov and A. Kusenko for discussions. 
This work has been supported by DGICYT under Grants PB95-1077
and SAB94-0325, by a joint CICYT-INFN grant, and by the TMR 
network ERBFMRXCT960090 of the European Union. H. N. has been 
supported by Ministerio de Educaci\'on y Ciencia. V. S. also 
got support from RFFR under grant N. 95-02-03724.

\vskip 1truecm

\newpage
\vglue 2.0cm
\centerline{\hskip -1cm
\psfig{file=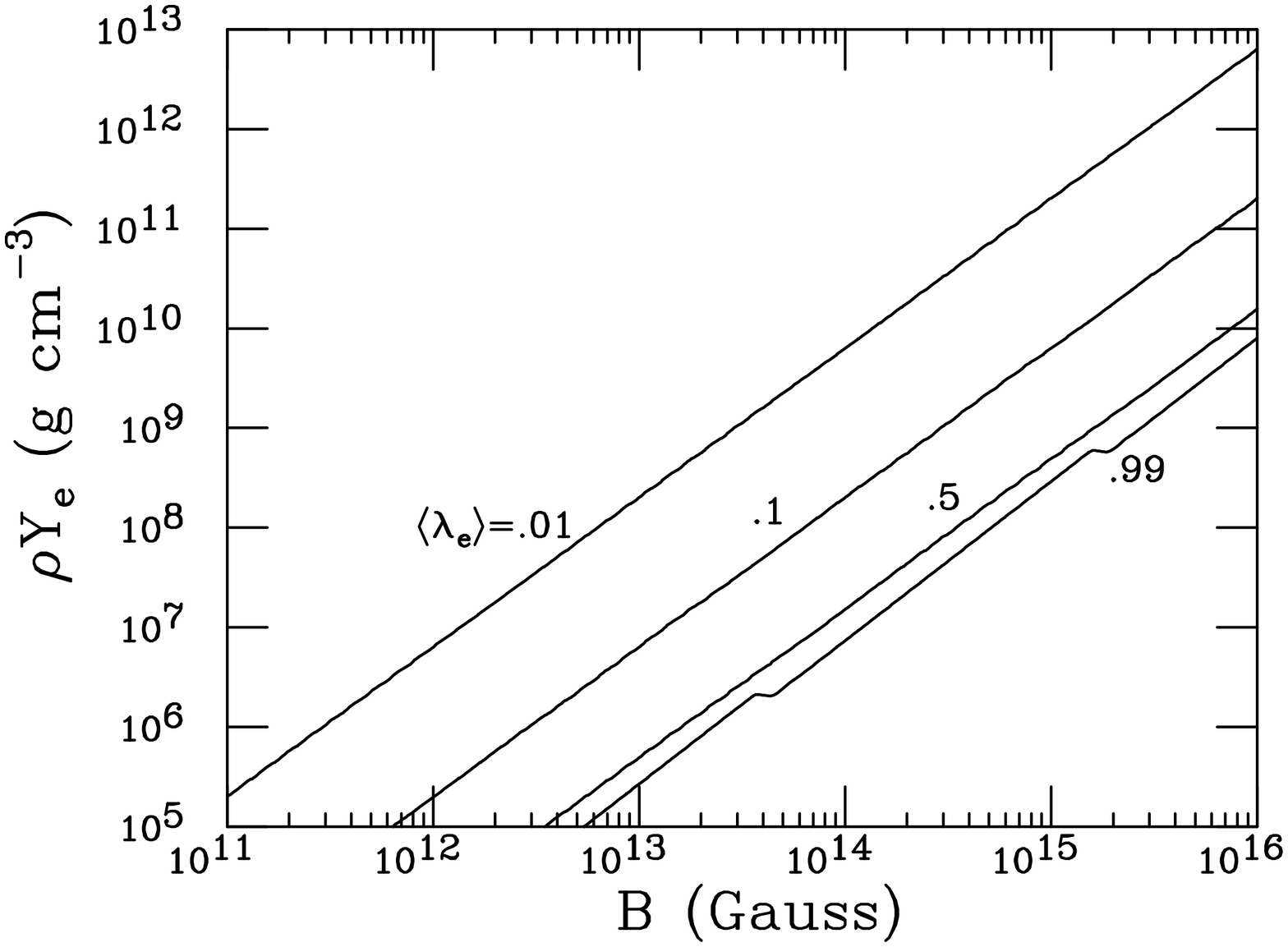,height=14.0cm,width=14.0cm,angle=90}}
\noindent
Fig. 1: Contour plot of the magnitude of electron polarization 
$\vev{\lambda_e}$ in the $\rho Y_e -B$ plane. We assumed the 
strong electron degeneracy and used the relation $p_F^3 = 3\pi^2 n_e$.
\vglue -0.6cm
\newpage
\vglue 1.0cm
\centerline{\hskip -1cm
\psfig{file=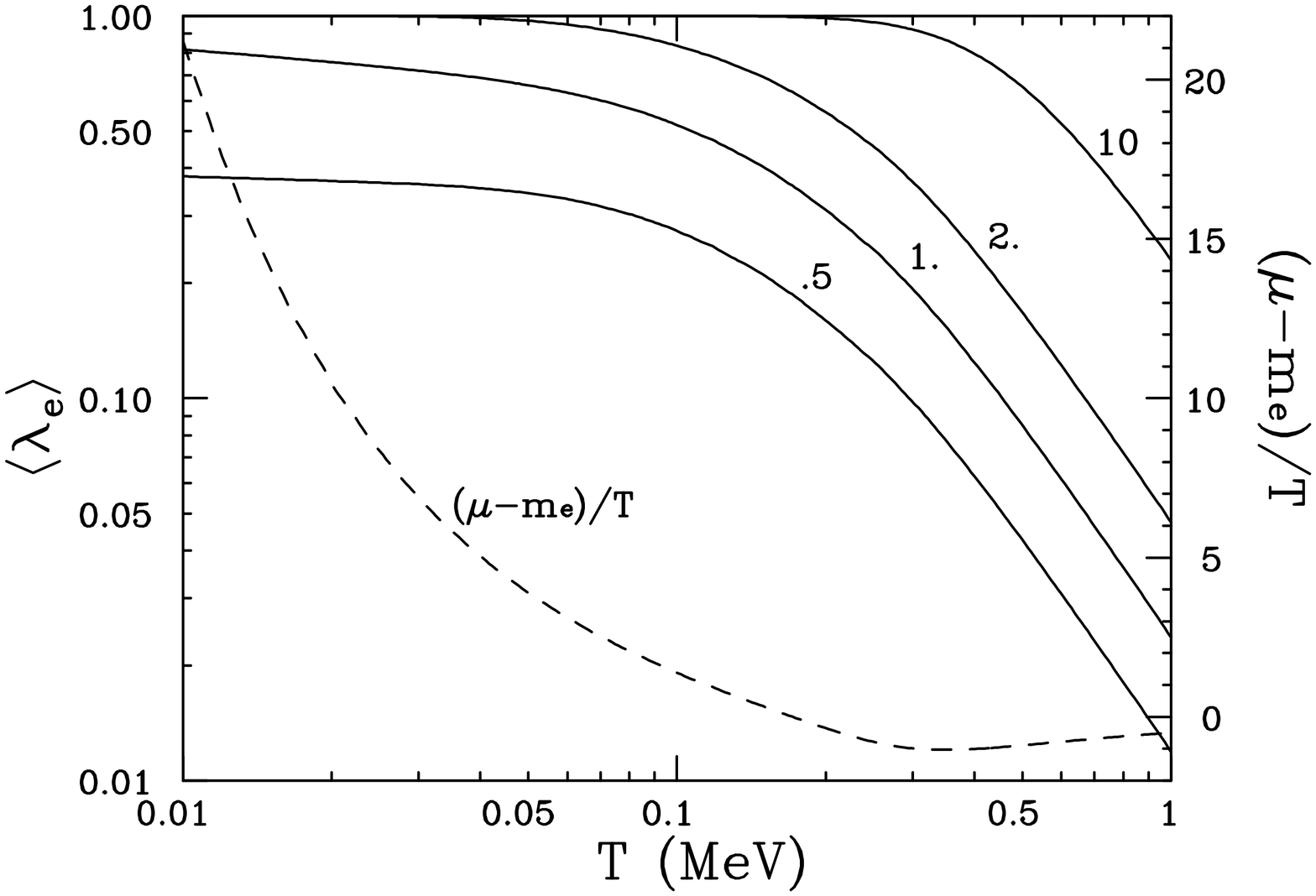,height=10.0cm,width=12.0cm,angle=90}}
\vglue -0.5cm
\noindent
Fig. 2a: Magnitude of electron polarization $\vev{\lambda_e}$ 
as a function of $T$ for different values of $2eB/p_F^2 = $ 
0.5, 1.0, 2.0, 10, indicated by numbers in the figure. 
We fixed the density, $\rho Y_e$, to be $10^6 (\mbox{g/cm}^3)$.
The degeneracy parameter $(\mu-m_e)/T$ is also 
plotted by dashed line. 
\vglue 0.6cm
\centerline{\hskip -1cm
\psfig{file=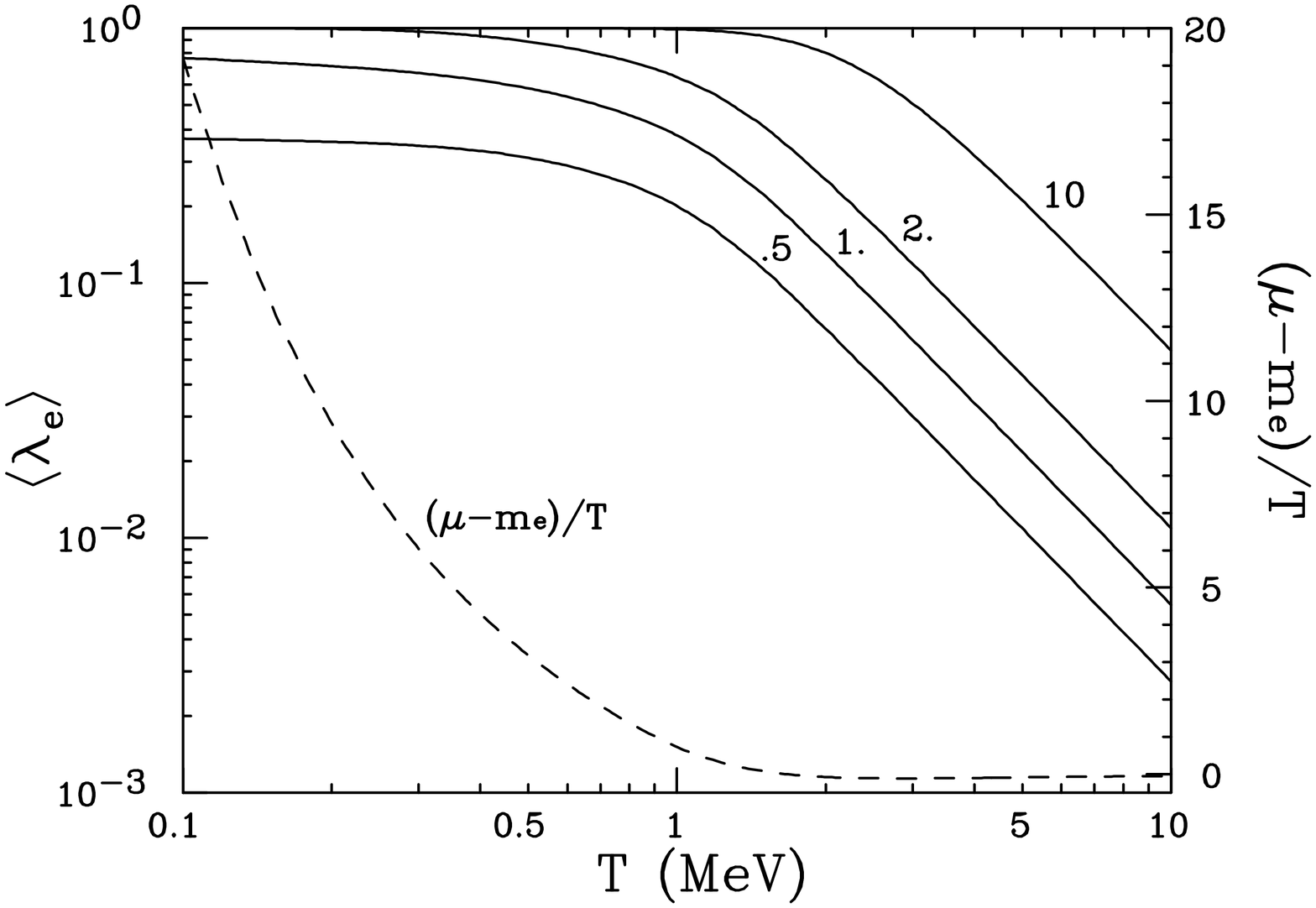,height=10.0cm,width=12.0cm,angle=90}}
\noindent
Fig. 2b: Same as the above figure but for 
$\rho Y_e = 10^8 (\mbox{g/cm}^3)$.
\newpage
\vglue 1.0cm
\centerline{\hskip -1cm
\psfig{file=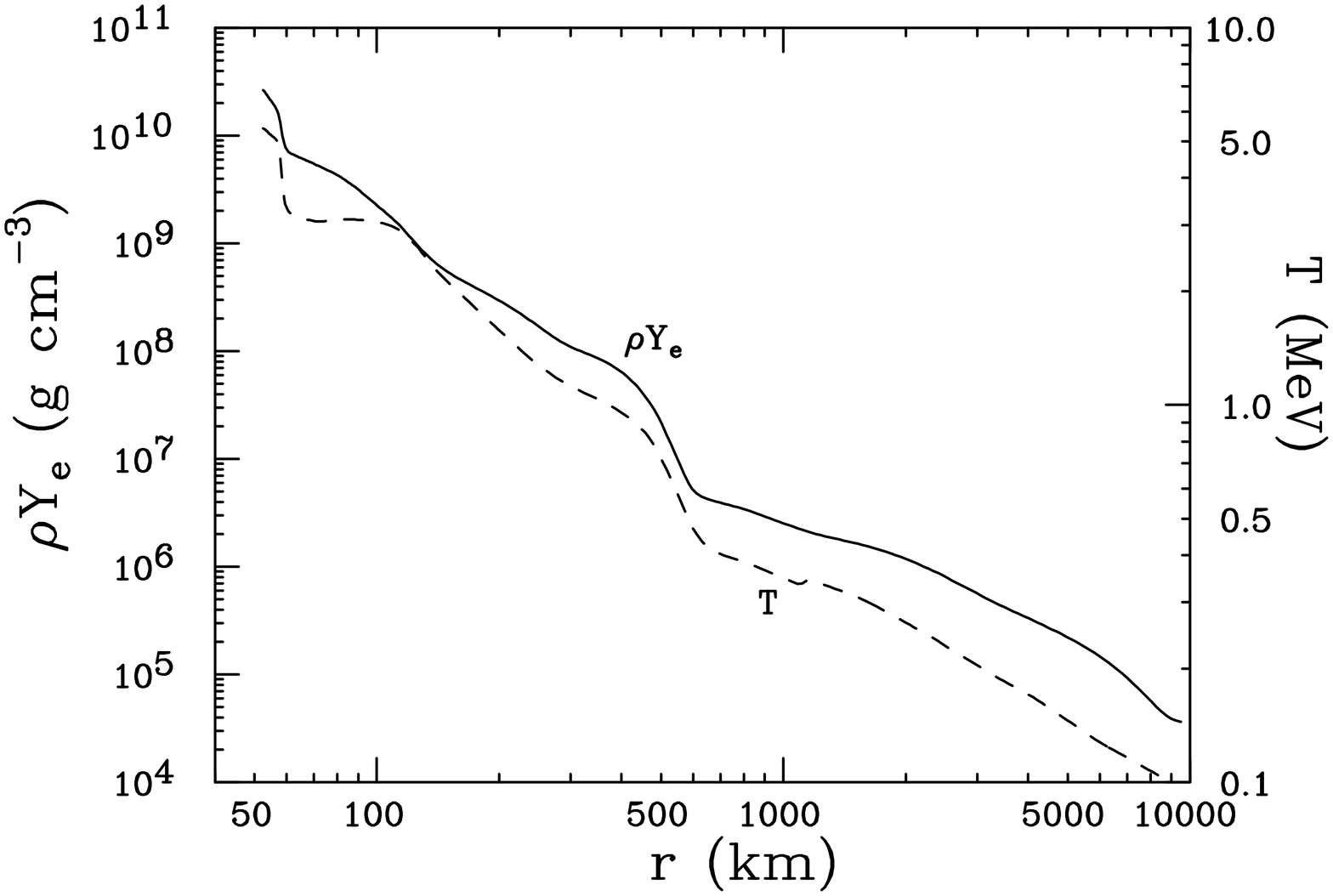,height=10.0cm,width=12.0cm,angle=90}}
\vglue -0.6cm
\noindent
Fig. 3a: The profiles of $\rho Y_e$ (solid line) and temperature 
(dashed line) 
at $t$ = 0.15 sec after the bounce from Wilson's model. 
\vglue 0.6cm
\noindent
\centerline{\hskip -1cm
\psfig{file=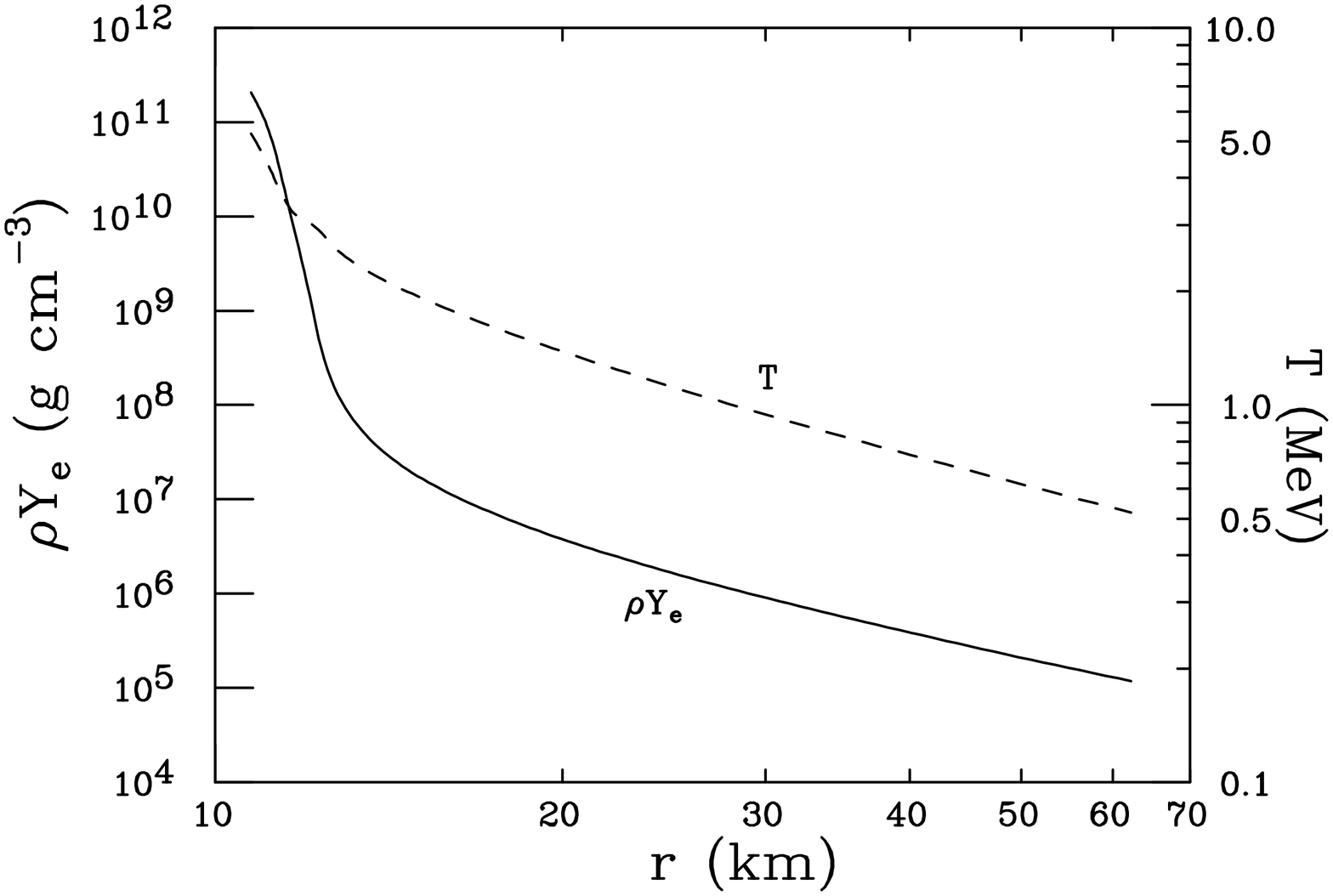,height=10.0cm,width=12.0cm,angle=90}}
\vglue -0.6cm
\noindent
Fig. 3b: Same as Fig. 3a but for the epoch at $ t \sim$  6 
sec after the bounce from Wilson's model. 
\newpage
\vglue 1.0cm
\centerline{\hskip -1cm
\psfig{file=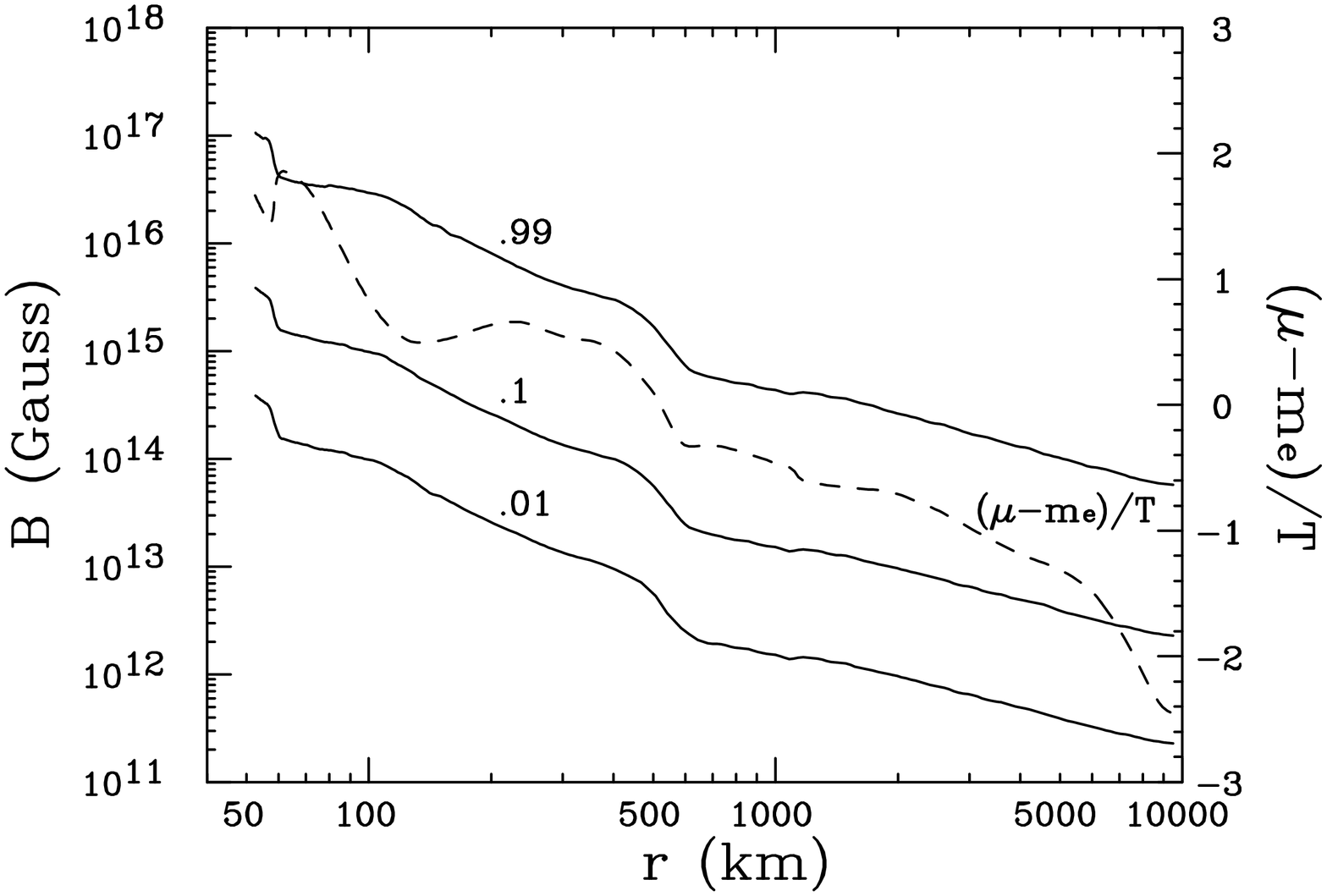,height=10.0cm,width=12.0cm,angle=90}}
\vglue -0.6cm
\noindent
Fig. 4a: Required magnetic field profiles for given polarization, 
$\vev{\lambda_e}$ = 0.01, 0.1 and 0.99 for density and temperature profiles
shown in Fig. 3a ($t=0.15$ sec after the bounce). 
The degeneracy parameter, $(\mu-m_e)/T$ is also plotted by 
dashed line.
\vglue 0.6cm
\centerline{\hskip -1cm
\psfig{file=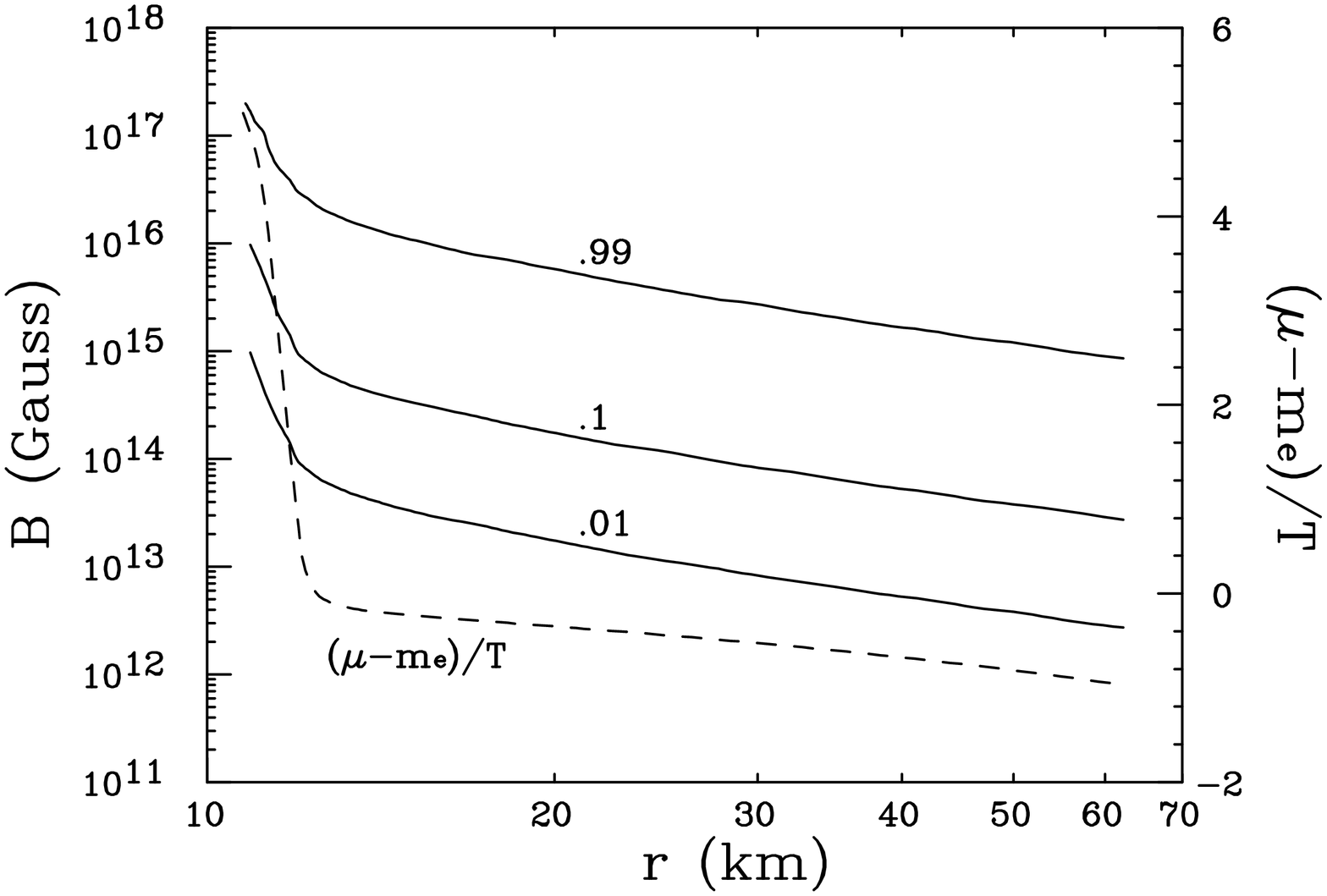,height=10.0cm,width=12.0cm,angle=90}}
\vglue -0.6cm
\noindent
Fig. 4b: Same as 4a but for later epoch ($t\sim 6$) sec 
of Wilson's model. 

\newpage
\vglue 2.0cm
\centerline{\hskip -1cm
\psfig{file=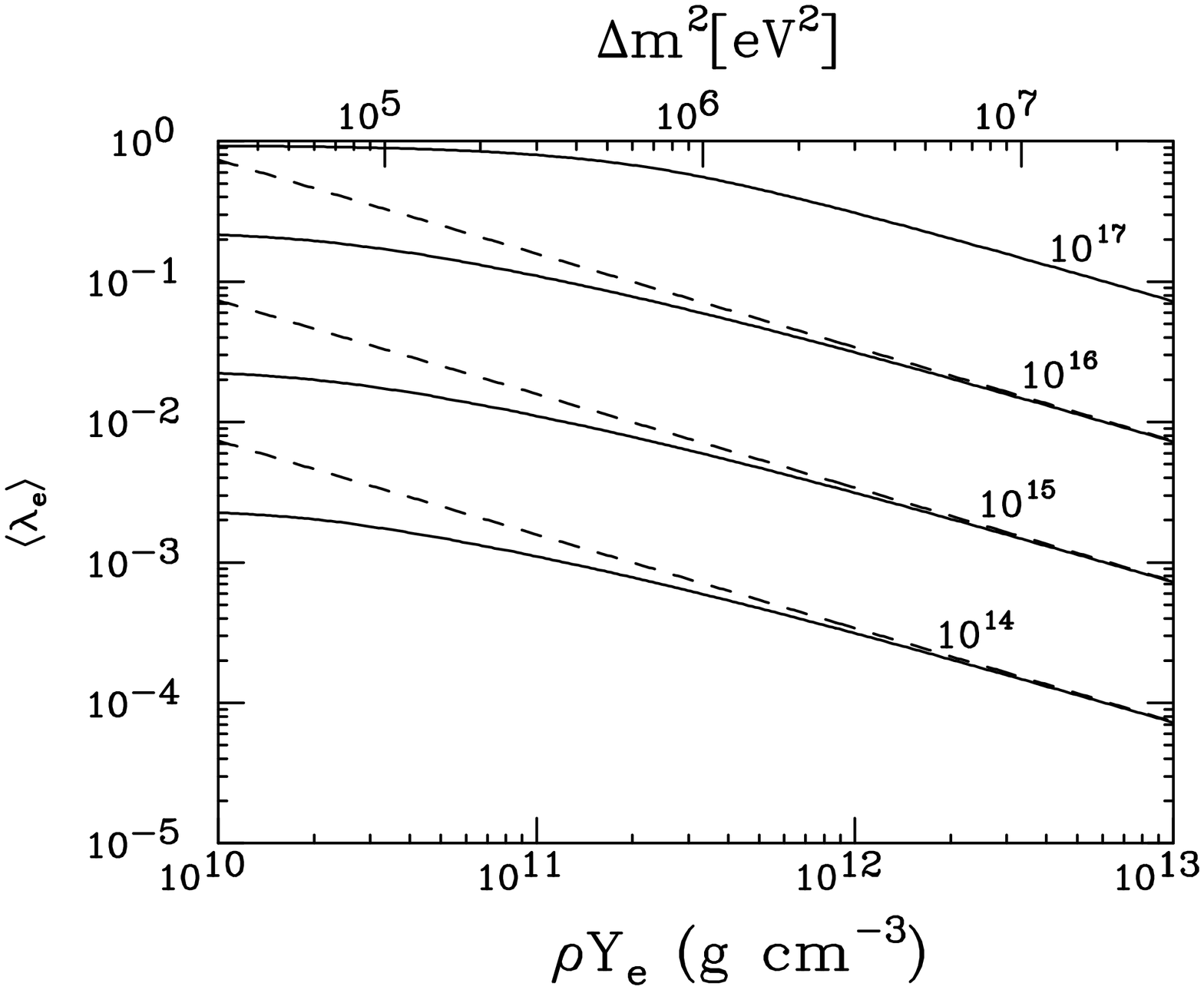,height=13.0cm,width=15.0cm,angle=90}}
\vglue -0.5cm
\noindent
Fig. 5: Magnitude of electron polarization $\vev{\lambda_e}$ 
as a function of $\rho Y_e$ for fixed temperature, $T=7$ MeV, 
and for different values of magnetic field strength, 
$B = 10^{14}, 10^{15}, 10^{16}$ and $10^{17} $ Gauss, 
indicated by the numbers in the figure. We also plot, except 
for $B = 10^{17}$ Gauss, $\vev{\lambda_e}$ calculated by 
eq.(40) by dashed lines. In the top, we indicate the 
corresponding $\Delta m^2$ values for which $E=20$ MeV neutrino 
undergo resonance for the case where no polarization effect exists. 

\end{document}